\documentclass[10pt]{article}

\usepackage{amsmath,amssymb,amsthm}
\usepackage{amsxtra}
\usepackage{amsfonts}
\usepackage{amssymb}
\usepackage{url}
\usepackage{hyperref}
\usepackage{esint}
\usepackage{eucal}
\usepackage{mathrsfs}
\usepackage{cite}
\usepackage{graphicx,graphics}

\newcommand\eq[1] {(\ref{#1})}

\newcommand{\nonum}{\nonumber \\}
\newcommand{\beqa}{\begin{eqnarray}}
\newcommand{\eeqa}[1]{\label{#1}\end{eqnarray}}
\newcommand{\beq}{\begin{equation}}
\newcommand{\eeq}[1]{\label{#1}\end{equation}}

\newcommand{\Real}{\mathop{\rm Re}\nolimits}

\newcommand{\Md}{\partial}

\newcommand{\Gb}{\beta}
\newcommand{\Gd}{\delta}

\newcommand{\Gve}{\varepsilon}

\newcommand{\Gk}{\kappa}

\newcommand{\Gl}{\lambda}

\newcommand{\Gm}{\mu}
\newcommand{\Gv}{\nu}

\newcommand{\Gt}{\theta}

\newcommand{\Go}{\omega}

\newcommand{\GO}{\Omega}

\newcommand{\bpm}{\begin{pmatrix}}
\newcommand{\epm}{\end{pmatrix}}

\def\Bx{{\bf x}}

\def\BD{{\bf D}}
\def\BE{{\bf E}}

\usepackage{graphics}
\usepackage{amssymb}

\title{A review of anomalous resonance, its associated cloaking, and superlensing}
\date{}
\begin{document}
\maketitle
\vskip -.5cm
\centerline{\large
  Ross C. McPhedran \footnote{School of Physics, The University of Sydney, Australia -- ross.mcphedran@sydney.edu.au}
  \,\, and \,\,
Graeme W. Milton \footnote{Department of Mathematics, University of Utah, USA -- milton@math.utah.edu,}}
\vskip 1.cm
\begin{abstract}
  We review a selected history of anomalous resonance, cloaking due to anomalous resonance, cloaking due to complementary media, and superlensing.    
\end{abstract}

\begin{center}{\bf Titre en Fran\c{c}aise}\end{center}
\begin{center}R\'{e}sonance anormale, et invisibilit\'{e} et super-resolution associ\'{e}es : un \'{e}tat de l'art\end{center}

\begin{center}{\bf Abstrait en Fran\c{c}aise}\end{center} 
\begin{center}Nous passons en revue une s\'{e}lection sur l'historique de la r\'{e}sonance anormale,
de l'invisibilit\'{e} associ\'{e}e \`{a} la r\'{e}sonance anormale, de l'invisibilit\'{e} associ\'{e}e aux milieux
compl\'{e}mentaires et de la super-r\'{e}solution\end{center} 
%%%%%%%%%%%%%%%%%%%%%%%%%%%%%%%%%%%%%%%%%%%%%%%%%%%%%%%%%%%%%%%%%%%%%%%%
\section{Introduction}
\setcounter{equation}{0}
%%%%%%%%%%%%%%%%%%%%%%%%%%%%%%%%%%%%%%%%%%%%%%%%%%%%%%%%%%%%%%%%%%%%%%%%%%%%%%%%%%%%%%%%%%%%%%%%%%%%%%%%%%%
In usual resonance as the loss goes to zero, one is approaching a pole of the associated linear response function. By contrast, anomalous localized resonance (ALR)
is associated with the approach to an essential singularity. The connection with essential
singularities is evident in figure 8 of \cite{Nicorovici:1993:TPT} and shown explicitly in the analysis in \cite{Ammari:2013:STN}, where the underlying theory was developed. 
Anomalous resonance has the following three defining features:

\begin{enumerate}
\item As the loss goes to zero, finer and finer scale oscillations develop as modes increasingly close to the essential singularity become excited.
\item As the loss goes to zero, the oscillations blow up to infinity in a region which is called the region of anomalous resonance, but outside of this
region the fields converge to a smooth field.
\item The boundary of the region of anomalous resonance depends on the source position.
\end{enumerate}

It is to be emphasized that the approach to an essential singularity does not necessarily imply anomalous resonance. In particular, anomalous resonance does
not occur for coated spheres, with the coating and the core each being isotropic and having constant complex dielectric constant
\cite{Ammari:2013:STN}. (It can occur if one allows for an anisotropic coating \cite{Ammari:2013:ALR}).
Also anomalous resonance, defined in this way, should be distinguished from the unusual feature that, in the presence of materials with negative
properties, corners or other singularities in the microstructure can behave like sinks of energy in the limit where the loss parameters of the materials tends to zero:
see page 378 of \cite{Milton:2002:TOC}, \cite{Qiu:2008:PLS}, section 2 of \cite{Helsing:2011:SSR} and \cite{Sihvola:2012:LLB,Estakhri:2013:PUB}.
Such behavior is associated with branch cuts.
For further mathematical development see \cite{BonnetBenDhia:2012:TIP,BonnetBenDhia:2013:RCN,BonnetBenDhia:2014:TDM,BonnetBenDhia:2014:TCS}. 

Here we will review anomalous resonance and its associated cloaking. We will also review the closely related topic of superlensing which would not be possible without anomalous resonance:
anomalous resonance provides the essential mechanism for a superlens producing an image of a point source beating the diffraction limit. At the same time it prevents the formation of a correct
image when a dielectric object being imaged is too close to the superlens. This is contrary to what one would believe from reading most of the literature on superlenses, since this limitation is rarely pointed out both in
published papers and in popular articles on the internet (such as in Wikipedia). It is unfortunate that wrong ideas may continue to propagate. This paper aims to contribute to correct the situation.  At this stage the literature is so vast that we can only cover, or even reference, a selected subset of papers, partly chosen for their importance and partly chosen because
we are familiar with them. 

Our analysis and the analysis in the papers that we
cite will mostly be for quasistatics. It is important to remember that quasistatics does not necessarily mean that the frequency of the applied field tends to zero.
A better procedure is that, at any fixed frequency $\Go_0$, the dimensions of the system should be shrunk to a size where they are much smaller than the free space wavelength
for the quasistatic approximation to be valid. 
%%%%%%%%%%%%%%%%%%%%%%%%%%%%%%%%%%%%%%%%%%%%%%%%%%%%%%%%%%%%%%%%%%%%%%%%
\section{The discovery of anomalous resonance and ghost sources}
\setcounter{equation}{0}
%%%%%%%%%%%%%%%%%%%%%%%%%%%%%%%%%%%%%%%%%%%%%%%%%%%%%%%%%%%%%%%%%%%%%%%%%%%%%%%%%%%%%%%%%%%%%%%%%%%%%%%%%%%
Back in 1993 we investigated with Nicolae Nicorovici the quasistatic effective properties of a square array of coated cylinders each having core dielectric constant $\Gve_c$ and
radius $r_c$, and with shell dielectric constant $\Gve_s$ and outer radius $r_s$, and embedded in a matrix having dielectric
constant $1$ \cite{Nicorovici:1993:TPT}. Surprisingly, we found that when $\Gve_s=-1$ the array had exactly the same effective dielectric constant as a square array of solid cylinders
having core dielectric constant $\Gve_c$ and radius $r_0=r_s^2/r_c$ embedded in a matrix having dielectric constant $1$. We subsequently looked
at a single coated cylinder, with the $z$-axis as its cylinder axis, in an infinite medium of dielectric constant $1$ subject to a non-uniform applied field at infinity
independent of $z$ \cite{Nicorovici:1994:ODP}. Again, when $\Gve_s=-1$, it was found that the effect of the shell was to magnify the core by a factor of $r_s^2/r_c^2$ so its response was equivalent to a solid cylinder
having core dielectric constant $\Gve_c$ and radius $r_0=r_s^2/r_c$. As the equivalent solid cylinder could have a very large radius when $r_c$ is small,
this marked the first discovery (in quasistatics) of what became known (for the full time-harmonic Maxwell's equations) as a superscatterer \cite{Yang:2008:SES}.
When $\Gve_c=1$ we observed that the coated cylinder
becomes invisible to any applied quasistatic field: in this sense, the shell cloaks the core. Inclusions that are invisible to any applied field at a prescribed frequency 
(not necessarily in the quasistatic regime) were also found by Dolin  \cite{Dolin:1961:PCT} as an example illustrating his discovery
of what is now known as transformation optics. Later, Alu and Engheta \cite{Alu:2005:ATP} found that coated spheres, built from appropriate materials, could be
invisible to incident plane waves at a prescribed frequency, thus extending the quasistatic results of Kerker \cite{Kerker:1975:IB}.
To illuminate our discoveries further
we investigated, in the same paper \cite{Nicorovici:1994:ODP}, the infinite body Green's function (fundamental solution) for a single coated cylinder in a matrix having dielectric
constant $1$ with a line dipole source at a distance $z_0$ from the cylinder axis. If the equivalence held, then when $\Gve_c\ne 1$ by the method of images the exterior field should be the same
as the field generated by the source line dipole and an image line dipole at the radius $r_0^2/z_0=r_s^4/(z_0r_c^2)$. This represents a paradox when $r_0^2/z_0>r_s$, i.e. when
$z_0<r_s^3/r_c^2$ (the latter being greater than $r_s$), as then the image line dipole lies outside the coated cylinder, i.e. there is a source there but we have not physically introduced
 such a source. To resolve this paradox we recognized that any material with a negative dielectric constant should also have some imaginary part due to resistive losses, and
therefore one should set $\Gve_s=-1+i\Gd$ and take the limit $\Gd\to 0$. Doing this we found that the field outside the radius $r_0^2/z_0$ converged to the field one expected, i.e
that due to the source dipole and (a ghost source) image dipole at the radius $r_0^2/z_0$. Inside the radius $r_0^2/z_0$ (and outside the coated cylinder) the field exhibited large
oscillations whose amplitude diverged and wavelength decreased as $\Gd\to 0$. This marked the first discovery of ghost sources and anomalous resonance: see Figure 1.
See also the unpublished introduction \cite{Milton:1994:URP} written prior to 1996. 
Anomalous resonance occurs when
as the loss goes to zero the field diverges in one region (the region of anomalous resonance) that is dependent on the position of the source, but converges to a smooth field 
outside this region. Insight into anomalous resonance can be obtained simply by considering the series expansion for a pole at the point $z=1$ in the complex $z=x+iy$-plane (not to be confused
with the $z$-axis):
\beq 1/(1-z)=1+z+z^2+z^3...\eeq{0.1}
If we truncate at high order the function on the right we obtain a polynomial that inside the radius of convergence $|z|<1$ almost looks like it has a singularity at $z=1$
(corresponding to a  ghost source) but outside the radius of convergence (corresponding to the region of anomalous resonance) exhibits enormous oscillations.
The difficulty is finding systems where the series truncation is somehow correlated with the loss in the system, as in the anomalously resonant coated cylinder system.

\begin{figure}[!ht]
\includegraphics[width=0.9\textwidth]{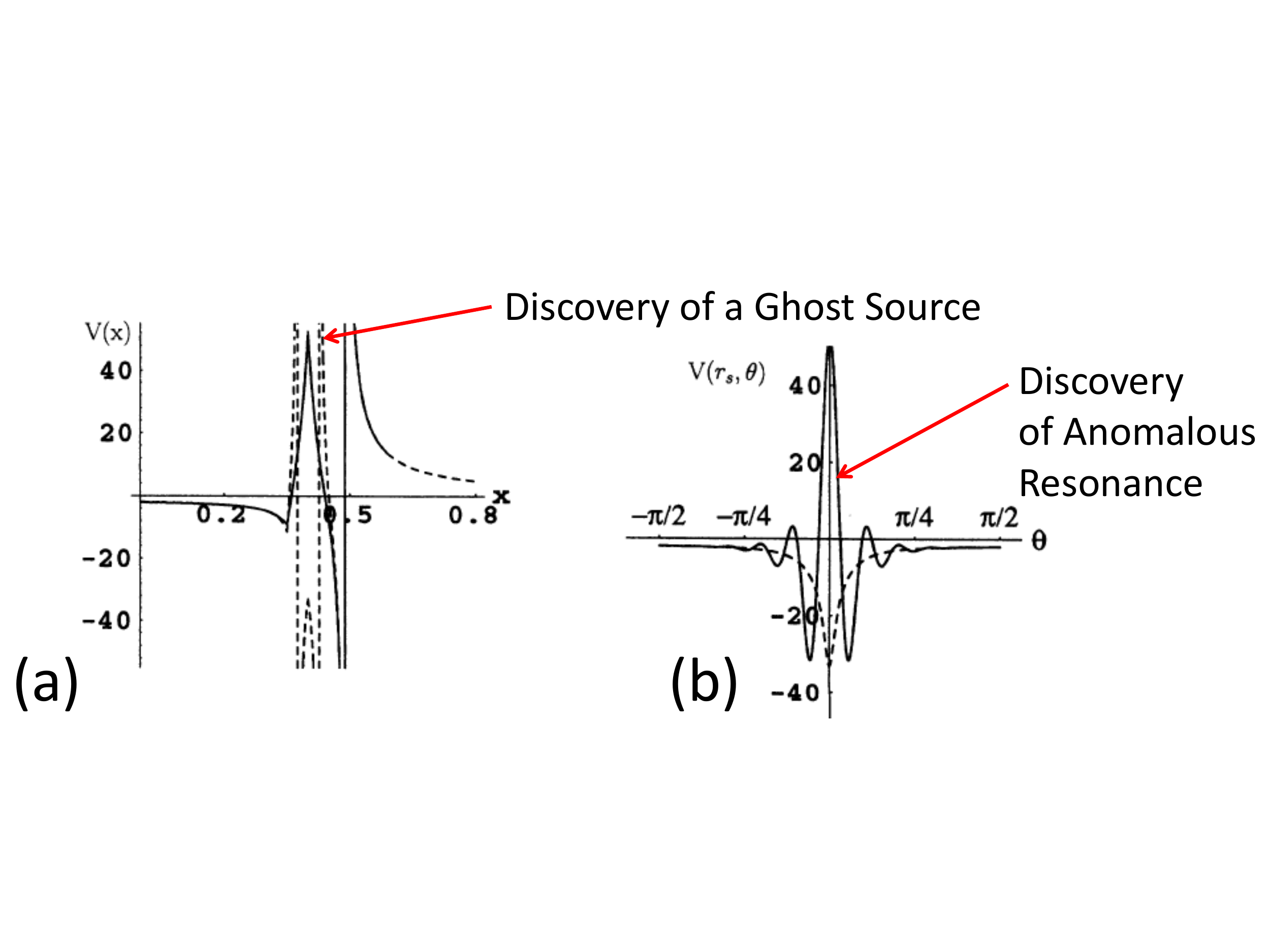}
\vskip -3.0cm
\caption{Reproduction, with permission, of Figure 2 of \protect\cite{Nicorovici:1994:ODP} highlighting our discoveries of ghost sources and anomalous resonance.
  The red arrows and accompanying text are new and are inserted to emphasize our findings. The solid curves show the actual potential, along the $x$-axis in (a) and
  around the outer surface of the coated cylinder, with parameters $\Gd=0.01$, $z_0=0.49$, $r_c=0.35$, $r_s=0.40$, $r_0=0.457$, and $r_0^2/z_0=0.426$. The latter radius,
  where the ghost source is located, is outside the coated cylinder. Outside this radius the solid curve in (a) slowly converges as $\Gd\to 0$ to the dashed curve that has a singularity
  at this radius, representing the ghost source. The proof of convergence was established based on the ratio test for the series expansion for the actual potential.
  The large oscillations of the potential in (b) clearly show the anomalous resonance. There were some mistakes in our initial
  analysis, but everything was correct for the case considered here of a source on the real axis.}
\end{figure}

\section{Anomalous resonance and ghost sources in superlenses}
\setcounter{equation}{0}
%%%%%%%%%%%%%%%%%%%%%%%%%%%%%%%%%%%%%%%%%%%%%%%%%%%%%%%%%%%%%%%%%%%%%%%%%%%%%%%%%%%%%%%%%%%%%%%%%%%%%%%%%%%
Anomalous resonance and ghost sources were rediscovered, both in numerical simulations and in theoretical works \cite{Rao:2003:AEW,Shvets:2003:ASP,Shvets:2003:PAM,Cummer:2003:SCS,Merlin:2004:ASA,Guenneau:2005:PCR,Podolskiy:2005:NSS,Podolskiy:2005:OSM}, that analyzed and provided the first sound basis for
Pendry's bold claim \cite{Pendry:2000:NRM} that the Veselago lens \cite{Veselago:1967:ESS}  consisting of a slab of thickness $d$ having dielectric constant $-1$ and relative magnetic permeability $-1$ would act as a superlens,
capable of breaking Abbe's diffraction limit and focusing light to arbitrarily small length scales. In fact, anomalous resonance and ghost sources provide the necessary mechanism
for superlensing. The papers \cite{Merlin:2004:ASA,Collin:2010:FDL,Gralak:2010:MME,Gralak:2012:NIM}
show that there is an essential singularity associated with this problem too.
With a point dipole source a distance $d_0<d$ from the lens, and with a slab having a dielectric constant of $-1+i\Gd$ and a relative magnetic permeability $-1+i\Gd$, in the
limit $\Gd\to 0$ two (possibly overlapping) anomalously resonant regions of width $2(d-d_0)$ develop around the two slab interfaces, and a ghost dipole source appears at the position of the expected image, at a point a distance $d-d_0$ from the slab, on the opposite side of slab from the source.
The wavelength of the oscillations in the anomalously resonant regions sets the length scale of resolution of the image ghost source. The connection with our earlier work on
the coated cylinder becomes clearer once one realizes that a slab can be regarded as a coated cylinder of infinite radius keeping $d=r_s-r_c$ fixed as $r_c\to\infty$. In this limit
our earlier analysis corresponds to a line dipole outside the Veselago lens in the quasistatic limit. This connection is made more explicit in the analysis of Section 4
of \cite{Milton:2005:PSQ}. Moreover, even at high frequencies where the free-space wavelength is comparable
or smaller than $d$, the fields in the anomalously resonant regions remain the same as in the quasistatic approximation because the field gradients are so large that the quasistatic
approximation remains valid there (see (4.25) and (4.26) and the discussion below them in \cite{Milton:2006:CEA} ). Our coated cylinder with $\Gve_s=-1$ and $\Gve_c=1$ became known as the perfect cylindrical lens \cite{Pendry:2003:PCL} or cylindrical superlens.

It is important to recognize that when $d_0<d$ the anomalously resonant regions occur around both interfaces of the slab. Their presence is crucial to energy conservation as when
$d_0<d/2$ they interact with the source and provide the ``radiation resistance'' needed to account for the energy flowing towards the regions of anomalous resonance which is dissipated
into heat there. When these anomalously resonant regions interact with the source they can destroy the claimed ``perfect imaging'' properties of the Veselago lens. On the other hand,
when $d>d_0>d/2$, then the image is in the region of anomalous resonance and it again can destroy the ``perfect imaging'' properties of the lens when the fields are measured at the plane
through the ghost source \cite{Milton:2005:PSQ}. In the presence of anomalously resonant fields acting on the source it is physically unlikely that the source will not react to these fields as
otherwise the energy
flowing to a source of fixed amplitude needs to be exactly tailored according to the magnitude of the anomalously resonant fields acting on the source. 

With a line dipole source outside the slab lens the electrical potential in its near vicinity takes the form,
\beq V=V_0+\frac{k_e\cos{\Gt}}{r}
-\frac{k_o\sin{\Gt}}{r}
-E_x r\cos{\Gt}
-E_y r\sin{\Gt}+O(r^2)
 \eeq{2.1}
where $V_0$ is an additive constant, $k_e$ and $k_o$ are possibly complex constants governing the amplitude of the dipole line source,
and $E_x$ and $E_y$ are the cartesian components of the field acting on the line source
(defined by \eq{2.1}), and $(r,\Gt)$ are polar coordinates around the line source. Then the quasistatic formula for energy conservation takes the form
(see  \cite{Milton:2005:PSQ}):
\beqa (\Go_0/2)\int_{\GO}\Gve''\BE\cdot\bar{\BE} & = & (\Go_0/2){\rm ~Imag}\negthinspace\int_{\GO}\BD\cdot\bar{\BE}
=(\Go_0/2){\rm ~Imag}\negthinspace\int_{\Md\GO}-\Gve\frac{\Md V}{\Md \underline{r}} \bar{V} 
\nonum
& = & \Go_0\pi{\rm ~Imag}[\bar{k}_e E_x-\bar{k}_o E_y]
\eeqa{2.2}
where the quantity on the left is the power generated per unit length of the line source, $\GO$ is a two-dimensional domain consisting of a cross-sectional plane
with an infinitesimal circle surrounding the source cut out from it, and the overline denotes complex conjugation. Notice that the energy flowing from (or to) the
source (or sink) is dependent on the values of $E_x$ and $E_y$. When the dipole source is in the region of anomalous resonance the fields $E_x$ and $E_y$
scale like $|\Gd|^{2(d_0/d)-1}|\log\Gd|$ which agrees with the scaling of the loss in the lens as demanded by \eq{2.2}. This blows up to infinity as $\Gd\to 0$ when
$d_0<d/2$.

\section{Cloaking due to anomalous resonance}
\setcounter{equation}{0}
In 2005 (private communication) Alexei Efros remarked that something was amiss in everyones understanding of the superlens. He had calculated the result just mentioned:
that when $d_0<d/2$ the electrical power consumed by the superlens (in the regions of anomalous resonance) with a constant amplitude source approaches infinity as $\Gd\to 0$.
He had thought that energy was not conserved
and that therefore the concept of the superlens was flawed. In a closer analysis we found (at least in the quasistatic limit) that \eq{2.2} shows energy is conserved. As $\Gd$ is reduced,
ever increasing amounts of power are drawn from the source due to its interaction with the anomalously resonant fields. The anomalously resonant fields provide a sort of ``optical molasses'' against which
the source has to work -- it is a type of radiation resistance. If one thinks of the source as being generated say by oscillating charges, then the forces generated
by the anomalously resonant fields acting on these charges are directed in constant opposition to their movement.

Later we realized that since any realistic dipole source, such as a source providing constant power or a polarizable dipole
(which becomes a source in the presence of an incident field) cannot provide ever increasing amounts of power, its amplitude must go to zero as $\Gd\to 0$. In this limit
it will become invisible outside the region of anomalous resonance - it should become cloaked.
We then proved this for an arbitrary finite number of polarizable line dipoles, or constant power dipole sources,
outside the quasistatic coated cylinder that lie within the cloaking region \cite{Milton:2006:CEA}. In particular,
as proved in section 3 of that paper,
if the total power produced by the dipole sources remains bounded, then the amplitude of each and every dipole in
the cloaking region must go to zero in the limit $\Gd\to 0$. With polarizable dipoles inside the cloaking region, the field acting on each of them must tend to zero
as  $\Gd\to 0$.
Our paper was perhaps the first to introduce the word ``cloaking'' into the scientific literature, outside
computer science. Shortly after the publication of our paper, the transformation based cloaking approaches of Pendry, Schurig, and Smith \cite{Pendry:2006:CEM} and Leonhardt \cite{Leonhardt:2006:OCM}
appeared, for the time harmonic Maxwell equation and
for geometric optics respectively: the former can be regarded as a combination of the transformation optics ideas for electromagnetism, which date back to Dolin \cite{Dolin:1961:PCT},
and singular cloaking transformations, which date back to Greenleaf, Lassas, and Uhlmann \cite{Greenleaf:2003:ACC}. One of the
interesting features of cloaking due to anomalous resonance, as opposed to transformation based cloaking, is that the cloaking region lies outside the cloak. Rather than guiding fields
around a collection of polarizable line dipoles in the cloaking region, the fields generated in the anomalously resonant regions are such as to almost cancel the fields acting on each
 polarizable line dipole. In the limit $\Gd\to 0$, the nodal lines of the total field amazingly arrange themselves so as to almost intersect all polarizable line dipoles in the cloaking region.  
 Numerical analysis \cite{Milton:2006:CEA,Nicorovici:2007:QCT} confirmed these predictions, and moreover showed that a line quadrupole in the cloaking region outside would be cloaked as well. The original paper
 also established cloaking of a polarizable line dipole outside the slab lens at all frequencies (not just in the quasistatic limit) and for a point dipole outside the three-dimensional
 slab lens in the quasistatic limit. In that paper it was suggested that ``it may be the case that any object of finite extent lying entirely within the cloaking region of the slab lens
 will be cloaked'' (in the limit as $\Gd\to 0$). While, for the coated cylinder, initial calculations of Bruno and Lintner \cite{Bruno:2007:SCS} suggested otherwise,
 a recent rigorous proof of Nguyen \cite{Nguyen:2017:CAO} has shown amazingly that cylindrical objects having a small but finite cross section that are near the coated
 cylinder are perfectly cloaked in the limit $\Gd\to 0$. This extended earlier work of \cite{Bouchitte:2010:CSO} that proved cloaking due to anomalous resonance of a cylinder
 with a radius going to zero as $\Gd\to 0$. 
 Even the paper of Bruno and Lintner established that small dielectric objects in the cloaking region can be partially cloaked in the limit as $\Gd\to 0$. The important
 conclusion is that the cylindrical superlens (and presumably also the slab superlens) does not properly image dielectric objects close to the lens even in the limit $\Gd\to 0$.

 As recognized by Leonhardt and Philbin \cite{Leonhardt:2006:GRE} the superlens can be obtained by using transformations to ``unfold'' a folded geometry having $\Gve'=\Gm'=1$
 everywhere. In two-dimensions the unfolding transformation is simply the inverse of the folding transformation,
 \beq x'=x,~y'=y~{\rm for}~~x<0;\quad x'=-x,~y'=y~{\rm for}~~0<x<d;\quad x'=x-2d,~y'=y~{\rm for}~~x>d. \eeq{3.0a}
 Using the rules of transformation optics this results in a material with
 \beq \Gve=\Gm=1~{\rm for}~~x<0~~{\rm and~for}~~x>d; \quad \Gve=\Gm=-1~{\rm for}~~0<x<d, \eeq{3.0b}
 which is the superlens. This is consistent with the mirroring property of each interface in a superlens \cite{Maystre:2004:PLM}.
 Anomalous resonance and cloaking also exist in other ``folded'' and equivalent ``unfolded geometries'' \cite{Milton:2008:SFG,Ammari:2013:ALR}.
 In these folded geometries (unlike in the folded geometry of Leonhardt and Philbin) it is important to keep the fields on the different ``sheets'' separate to analyze
 the anomalous resonance and associated cloaking. In the unfolded geometries the material in the shell generally has a position dependent anisotropic dielectric tensor field. 

 Anomalous resonance due to cloaking with a continuous source in the cloaking region (rather than a discrete set of dipoles) was first investigated in \cite{Ammari:2013:STN}.
 The magnitude of the source is scaled so the net time averaged average power dissipation,
 \beq  (\Go_0/2)\int_{\mathbb{R}^d}\Gve''\BE\cdot\bar{\BE}= \Gd(\Go_0/2)\int_{{\rm lens}}\BE\cdot\bar{\BE} \eeq{3.1}
 remains constant as $\Gd\to 0$. Then cloaking due to anomalous resonance is said to occur if the field becomes
 localized (in the region of anomalous resonance) and converges to zero outside of it. That paper also introduced the concept of weak cloaking due to anomalous
 resonance (weak CLAR). In that scenario cloaking due to anomalous resonance occurs for a sequence of values of $\Gd$ tending to zero, but not necessarily for all sequences of
 values of $\Gd$ tending to zero (which would imply strong CLAR). 

 One may ask if cloaking due to anomalous resonance applies to coated bodies of shapes that are not cylindrical.
 Then the separation of variables  method in \cite{Milton:2005:PSQ,Milton:2006:CEA,Ammari:2013:STN}
 is not appropriate. Using a variational approach Kohn, Lu, Schweizer and Weinstein \cite{Kohn:2012:VPC} establish that quasistatic cloaking due to anomalous resonance
 occurs with a variety of cylindrical geometries that have a non-circular inner core, but a circular outer boundary. Kettunen, Lassas and Ola \cite{Kettunen:2018:AEA}
 went beyond the quasistatic limit and studied anomalous resonance and its absence in a variety of shaped bodies containing isotropic material in two and higher dimensions.

 One may also ask if anomalous resonance and cloaking due to anomalous resonance occurs in other physical equations. Using a direct mathematical exact equivalence between
 the complex quasistatic equations and certain magnetoelectric or thermoelectric equations (see section 6 in \cite{Milton:2005:PSQ}) it follows that
 anomalous resonance and the associated
 cloaking occurs in these systems of equations as they loose ellipticity in an appropriate way. Also anomalous resonance and cloaking due to anomalous resonance
 has been shown to occur in the quasistatic elasticity equations \cite{Li:2016:ALR,Ando:2017:SNP,Ando:2018:SPN}. In these papers they find that this occurs
 at the essential singularity of the relevant Neumann-Poincar\'e operator. Assuming the Lame moduli of the two phases to be $\Gl_1$, $\Gm_1$ and $\Gl_2=c\Gl_1$, $\Gm_2=c\Gm_2$
 they find this occurs in two-dimensions when
 \beq c=-\frac{\Gl_1+3\Gm_1}{\Gl_1+\Gm_1}=-\frac{\Gk_1+2\Gm_1}{\Gk_1},\quad\text{or}\quad c=-\frac{\Gl_1+\Gm_1}{\Gl_1+3\Gm_1}=-\frac{\Gk_1}{\Gk_1+2\Gm_1}, \eeq{3.2}
 where $\Gk_1=\Gl_1+\Gm_1$ is the bulk modulus of phase 1. Let $\Gk_2=\Gl_2+\Gm_2=c\Gk_1$ be the bulk modulus of phase 2.
 Now in any simply connected region that is devoid of a sources
 a stress field that solves the elasticity equations will also solve the elasticity equations if a constant is added to the inverse shear moduli, and the same
 constant subtracted from the inverse bulk moduli (see section 4.5 of \cite{Milton:2002:TOC} and references therein).
 This strongly suggests that one can remove the constraint that the Lame moduli of
 phase 2 are a multiple $c$ of the Lame moduli of phase 1, and that more generally anomalous resonance and cloaking will occur in these two-dimensional systems when
 \beq\Gv_1E_2-\Gv_2E_1=3E_2+E_1\quad\text{or}\quad\Gv_2E_1-\Gv_1E_2=3E_1+E_2,
 \eeq{3.3}
 in which
 \beq \Gv_1=\frac{\Gk_1-\Gm_1}{\Gk_1+\Gm_1}, \quad \Gv_2=\frac{\Gk_2-\Gm_2}{\Gk_2+\Gm_2}, \quad E_1=\frac{4\Gk_1\Gm_1}{\Gk_1+\Gm_1}, \quad E_2=\frac{4\Gk_2\Gm_2}{\Gk_2+\Gm_2}
 \eeq{3.4}
 are the Poisson's ratios, and Young's moduli of the two phases. Anomalous resonance and cloaking have also been shown to occur for elastodynamics, without making
 a quasistatic approximation \cite{Deng:2019:SPN}.

 We remark that transformation based cloaking associated with many other physical equations has also been extensively studied and experimentally observed. The list
 is too long to include here, especially as transformation based cloaking is not the focus of this review.

 \section{A closer analysis of the lossless perfect lens and in the limit as the dispersion goes to zero}

 For a long while the hope persisted that the Veselago lens would be perfect if the lens was truly lossless. However, then there is no solution to the time harmonic equations with
 a dipole energy source a distance $d_0<d$ from the lens  unless one inserts a energy sink at the image point in the lens and an energy source at the image point outside the lens:
 each interface of the lens mirrors the field (or its extension) \cite{Maystre:2004:PLM} and thus mirrors the field singularity.
 Each mirrored field singularity is interpreted as a sink or source depending on whether there is a net flow of energy towards it or away from it: see the numerical
 simulations in \cite{Rosenblatt:2017:PDE}. Physically inserting such sources or sinks assumes prior knowledge of the source and thus clearly diminishes the utility of the lens, so we disregard this possibility. 

As discussed in \cite{Milton:2006:OPL} insight into the behavior of the lossless Veselago lens can be obtained
 by taking a source that is turned on at some time. From equation (62) in \cite{Yaghjian:2006:PWS} it is seen that a
source of constant strength $E_0$ switched on at $t=0$ creates an electric
field which near the back interface (and outside the lens) scales approximately as
\beq E \sim E_0 t^{1-d_0/d}.
\eeq{4.a}
The stored electrical energy $S_E(t)$ will scale as the square of this, and consequently
the time derivative of the stored electrical energy will scale approximately as
\beq \frac{dS_E}{dt}\sim E_0^2t^{1-2d_0/d},
\eeq{4.b}
which blows up to infinity as $t\to \infty$ when $d_0<d/2$. If the source produces a
bounded amount of energy per unit time we have a contradiction. The
conclusion is that if the energy production rate of the source is
bounded then necessarily the amplitude $E_0$ must decrease to zero as $t\to \infty$.

More can be said \cite{Milton:2006:OPL} if we take a source that is turned on exponentially slowly, i.e. with a time dependence $e^{t/T}e^{-i\Go_0 t}=e^{-i\Go t}$ corresponding to a complex frequency $\Go=\Go_0+i/T$ where $T$
 is a measure of the time the source has been ``switched on'' prior to time $t=0$. (At times say before $t=-10T$ the source amplitude in negligibly small while for times between $t=-T/10$ and $t=0$
 is approximately constant). For simplicity we only analyze the quasistatic case, but similar conclusions should hold when one considers the full Maxwell equations. 

 Around the frequency $\Go_0$ the dielectric constant in the shell has an expansion
 \beq \Gve_s=-1+(\Go-\Go_0)a+O((\Go-\Go_0)^2)=-1+ia/T+ O(1/T^2), \quad {\rm where}~a=\frac{d\Gve_s(\Go)}{d\Go}{\Huge{|}}_{\Go=\Go_0}, \eeq{4.1}
 and $a$ is the dispersion at the frequency $\Go_0$.
 Thus at long times $T$ the mathematical analysis is the same as in the time harmonic case with the dielectric constant of the shell having a
 very small imaginary part $\Gd\approx a/T$. A correspondence of this sort was noted before \cite{Yaghjian:2006:PWS}
but not fully exploited.
 Crucially, the anomalously resonant region in front of the lens still persists and again causes cloaking. After the source has been ``switched on'', say between times
 $t=-T/10$ and $t=0$, the fields are very nearly time-harmonic. With $a>0$ the electrical power produced by the source, say averaged over this time period, will be again given by the
 right hand side of \eq{2.2} and for fixed source amplitudes $k_e$ and $k_o$ will scale like
 \beq |\Gd|^{2(d_0/d)-1}|\log\Gd|\approx |a/T|^{2(d_0/d)-1}|\log(a/T)| \eeq{4.1a}
 and diverge as $T\to \infty$. If we want to avoid this power divergence then we need to rescale the source amplitudes $k_e$ and $k_o$ by the reciprocal of the quantity
 in \eq{4.1a}. Then $k_e$ and $k_o$ will go to zero as $T\to\infty$, i.e. the source will be cloaked as $T\to\infty$. Thus we arrive at a scenario where the source fades
 from view, both when it is viewed from behind and in front of the lens, as the time ``T'' during which it is switched on is increased: essentially all of its energy
 is drawn to build up the fields in the regions of anomalous resonance \cite{Milton:2006:OPL}. (A somewhat analogous effect occurs with band-limited superresolution
 \cite{DiFrancia:1952:NEEP}
 where, as the width of the focal spot is decreased, increasingly more energy is necessarily diverted to the side lobes and, correspondingly, the relative amplitude
 at the image spot necessarily decreases \cite{Shim:2019:MFS}.) With this scaling, the fields in the slab lens become localized to within the regions of
 anomalous resonance as $T\to\infty$. When $d>d_0>d/2$, then the image is in the region of anomalous resonance and it again can destroy the ``perfect imaging''
 properties of the lens. The interference
 of the surface waves associated with anomalous resonance and the image was also concluded by Collin \cite{Collin:2010:FDL} in a more complete analysis. 

 If we let $\BE(\Bx)$ denote the complex field that solves the time harmonic equations with $\Gve_s=-1+ia/T$, then the physical electric field
  that solves the equations in the lossless perfect lens should at large times $T$ be approximately
  \beq \widetilde{\BE}(\Bx,t)=\Real[e^{i(\Go_0+i/T)t}\BE(\Bx)]=e^{-t/T}\Real[e^{i\Go_0t}\BE(\Bx)].  \eeq{4.2}
%While there is no loss either in the slab or surrounding it, there is a buildup of electrical energy density $W(\Bx,t)$ given by the Brillouin formula. In the slab   
%\beq W(\Bx,t)=W_s(\Bx,t)\equiv\frac{d}{d\Go}(\Go\Gve_s(\Go))e^{-2t/T}|\BE(\Bx)|^2\approx e^{-2t/T}[-1+a\Go_0]|\BE(\Bx)|^2.
%\eeq{4.3}
%Thus it will appear that there is an absorption of energy in the slab (actually going into the field) given by
%beq P_s(\Bx,t)=\frac{d}{dt}W_s(\Bx,t)\approx 2e^{2t/T}[-1+a\Go_0]|\BE(\Bx)|^2/T, \eeq{4.5}
%hich will be negative when $a<1/\Go_0$, then requiring an active material in the slab to produce this energy. 
%utside the lens, 
%beq W(\Bx,t)=W_o(\Bx,t)\equiv |\widetilde{\BE}(\Bx,t)|^2=

 We consider four cases:

 (i) $a>1/\Go_0$ so that the local electric field energy density in the slab is positive. At the frequency $\Go_0$ this energy density is given by the Brillouin energy:
 \beq W_s(\Bx,t)\approx \frac{d}{d\Go}(\Go\Gve_s(\Go))|\widetilde{\BE}(\Bx,t)|^2 \approx [-1+a\Go_0]|\widetilde{\BE}(\Bx,t)|^2\approx e^{2t/T}[-1+a\Go_0]|\BE(\Bx)|^2\eeq{4.4}
 In fact if the material in the slab is passive one has the tighter inequality that $a>4/\Go_0$ \cite{Yaghjian:2005:IBA,Yaghjian:2006:PWS,Milton:2006:OPL}. Note that while there is no loss either in the slab or surrounding it,
 there is a buildup of the electrical energy density $W_s(\Bx)$ given by the Brillouin formula. Thus it will appear that there is an absorption of energy given by
 \beq P_s(\Bx,t)=\frac{d}{dt}W_s(\Bx,t)=2e^{2t/T}[-1+a\Go_0]|\BE(\Bx)|^2/T \eeq{4.5}

 (ii) $1/\Go_0>a>0$ so that the local electrical field energy density in the slab, given by the Brillouin formula, is negative, which requires the slab to be an active material

 (iii) $a=0$ so that there is absolutely no dispersion, which again requires an active material in the slab.

 (iv) $a<0$ so that the dispersion is negative and once again requires an active material in the slab.

 The case (i) was studied in \cite{Milton:2006:OPL}, and as discussed above the source becomes cloaked as $T\to\infty$.
 In case (ii) for a given $T$ the imaginary part of $\Gve_s=-1+ia/T$ is lower than in case (i) which means that cloaking occurs quicker than in case (i):
 if we halve $a$ and at the same time halve $T$ then the solution remains unchanged. Now power from the source and power from the fields in the active material in the
 slab lens both flow towards the regions of anomalous resonance outside the slab, both in front and behind the slab. There should also be a flow of energy outwards
 along the slab interfaces carried by the surface plasmons associated with the anomalous resonance, but we have not investigated this.
 In case (iii) we would need to look at higher order terms in the expansion \eq{4.1} in powers
 of $1/T$ to determine the asymptotic behavior as $T\to\infty$.
In case (iv) $\Gve_s=-1+ia/T$ has a negative imaginary part and the solution is obtained by taking the
 complex conjugate of the solution with $a>0$. In particular this means
 that the ``so-called'' dipole source that one puts in front of the lens needs to be a sink of energy, trapping the energy that is flowing from the slab of active material
 that forms the ``lens''.

 Obviously these considerations show that the response of a lossless superlens is not at all quick, rather it takes a long while for the lens  to achieve deep subwavelength
 resolution.

 Since in the time domain, the time $T$ the source has been ``on'' is mathematically equivalent to having a loss parameter $\Gd=a/T$ a curious effect may happen if
 a source causes weak CALR and not strong CALR. Then one may expect the source (normalized so that it provides constant power) to flash on and off in brightness as time progresses
 as seen from either in front of the lens or behind the quasistatic slab lens.

 More numerical work and rigorous theoretical work needs to be done in studying the quasistatic lens and the Veselago lens in the time domain, especially for sources that are
 turned on at a specific time and which supply a constant power (averaged over a cycle of oscillation) thereafter.  Collin \cite{Collin:2010:FDL} studied
 the interesting case of what happens with a ``Drude type metamaterial'' when a time harmonic source is turned on and then turned off at a later time. 
 Further studies were made in the papers \cite{Gralak:2010:MME,Gralak:2012:NIM,Cassier:2017:STM} where the response due to a time harmonic point source
 turned on in front of a half-space again containing a ``Drude type metamaterial''
 was investigated. Anomalous resonance does not occur with such a half space, rather the fields diverge linearly with time everywhere and consequently the image of a point
 source is not a point source even in the limit as $t\to\infty$ as shown by Gralak and Maystre \cite{Gralak:2012:NIM}: the resonant fields shroud the image. 
 All these analyses are for constant amplitude sources, rather than constant power sources.

 \section{Sensitivity of Anomalous Resonance, Cloaking, and Superlensing}

 As anomalous resonance is associated with essential singularities and as the behavior of analytic functions is quite wild around essential singularities it should
 come as no surprise that anomalous resonance, and hence cloaking due to anomalous resonance and superlensing, is quite sensitive to the material moduli. An indication of
 this sensitivity is that the cloaking region for the coated cylinder which has a radius $\sqrt{r_s^3/r_c}$ when $\Gve_c= 1$ changes dramatically to the radius
 $r_s^3/r_c^2$ when $\Gve_c\ne 1$ \cite{Milton:2006:CEA}. In fact, by considering the quasistatic slab superlens of thickness $d$ with $\Gve_s=-1+i\Gd$
 being the dielectric constant of the slab and with dielectric constants in front and behind the superlens of $1$ and $\Gve_c= 1+i(\Gd+\Gl\Gd^\Gb)$, respectively,
 with $\Gl$ and $\Gb$ being real constants satisfying $1>\Gb>0$, one finds that the position of the cloaking region depends on the value of $\Gb$ \cite{Meklachi:2016:SAL}.
 Specifically, as follows from equation (4.11) in that paper, the cloaking region extends a distance $d/(\Gb+1)$ in front of the slab. This varies continuously
 between $d$ and $d/2$ as $\Gb$ varies between $0$ and $1$. Early studies of superlenses also show sensitivity to the material moduli of the lens: see, for example,
 the excellent paper of Merlin \cite{Merlin:2004:ASA}.

 Interestingly, Xiao, Huang, Liu, and Chan \cite{Xiao:2015:EPC} show that while a small dielectric cylinder near the slab lens can be cloaked due to anomalous resonance,
 at nearby frequencies the opposite can also occur: it can become more visible due to a cylinder-slab resonance. 

 The sensitivity to material moduli is reduced as the loss parameter $\Gd$ is increased, but increasing $\Gd$ greatly reduces the resolution of the superlens. 

 \section{Cloaking due to complementary media}

 A different sort of superlens related cloaking, called cloaking due to complementary media, was
 proposed by Lai, Chen, Zhang, and Chan \cite{Lai:2009:CMI}. The idea, which has its origin in work by Pendry and Ramakrishna (see Figure 2 in \cite{Pendry:2003:FLN}),
 is that to cloak a given dielectric object close to the lens one should insert an appropriate
 cancelling ``anti-object'' in the lens. For the quasistatic slab lens if the front interface is at $x=0$ and the object lying within a distance $d$ has a dielectric constant
 $\Gve(x,y,z)$ then inside the lens one should modify the dielectric constant to $\Gve(x,y,z)=-\Gve(-x,y,z)$. If one takes the field for $x<-d$ without the object, antiobject and lens being present, and then analytically extends it to $x<0$ in the presence of the object, then the mirroring property allows us to reconstruct the field everywhere
 without disturbing the field for $x<-d$. Cloaking due to complementary media also occurs with bianisotropic media \cite{Liu:2014:FFE}.
 
 Some caution is needed since in the numerical simulations in Figure 2 of \cite{Lai:2009:CMI} and Figures 8 and 10 of \cite{Liu:2014:FFE}
 it looks like regions of anomalous resonance are appearing, and
 anomalous resonance does not enter into the argument used to justify the cloaking. Also we know that the argument of complementary media is sometimes flawed. Specifically,
 the field generally will not have an analytic extension from $x<-d$ to $x=0$ with the object being present. (The field outside the object generally will have
 singularities inside the object if one analytically extends it to the inside of the object while keeping $\Gve=1$ everywhere. These singularities will not
 generally be in the extension of the field from $x<-d$ to $x<0$, while keeping $\Gve=1$ everywhere, as required by the principle of complementary media.)
 By the argument  of complementary media the quasistatic slab lens is cancelled by a slab of material of moduli $\Gm=1$ directly behind it having thickness $d$.
 But this would imply the perfect
 imaging of polarizable dipoles, constant power sources, and dielectric objects close to the front of the lens, and we know that these are not perfectly imaged
 (even in the limit $\Gd\to 0$). Thus it appeared that the foundations of cloaking due to complementary media were on rather shaky grounds. Fortunately, in major
 advances, Nguyen \cite{Nguyen:2016:CCM} and Nguyen and Nguyen \cite{Nguyen:2015:CAL1} rigorously proved cloaking due to complementary media, subject to certain assumptions. 
 
 Associated with cloaking due to complementary media is illusion optics \cite{Lai:2009:IOO} and these simulations also appear to show anomalous resonance.

\section{Classifying different types of cloaking}
One way of classifying cloaking techniques is whether they satisfy the {\em ostrich effect} \cite{McPhedran:2009:CPR}. This term was designed to 
highlight whether the technique had the undesirable property wherein {\em the large object hides the small object, but the large object does not hide itself}.
Cloaking by anomalous resonance passes this ostrich test when $\Gve_c=1$, but not when $\Gve_c\ne 1$ as then the coated cylinder is visible as a larger cylinder of radius $r_s^2/r_c$
and dielectric constant $\Gve_c$. Classical single frequency cloaking by transformation optics also passes it: the cloaking system as well as the object to be hidden are difficult or impossible to detect.
% I would disagree with your comment about the carpet cloak. It is what sits on the ground (like an ostrich) and does not include the ground itself.
%However, the technique of carpet cloaking (references)  and variants in the literature do not, in that a cloaking region is created around a system which is open to detection. Whether this is a crucial defect would depe%nd on the requisites of the problem for which cloaking was put forward as a solution.

Another way of classifying cloaking techniques is whether the object to be cloaked is outside the cloaking device or not. This is a feature of cloaking due to anomalous resonance,
and cloaking due to complementary media,  and is called
exterior cloaking as opposed to the interior cloaking associated with transformation based cloaks. When it boils down to it, the exterior cloaking of anomalous resonance is due to polarization
charge sources at the surface of the lens. This then motivates one to investigate active exterior cloaks, where sources are chosen to achieve exterior cloaking. This was successfully done by
choosing sources to create a {\em quiet zone}, where the fields are small and the object to be cloaked can be hidden, while only slightly disturbing the fields outside a certain radius
\cite{Vasquez:2009:AEC,Vasquez:2009:BEC,Vasquez:2011:ECA,Vasquez:2012:MAT,Norris:2014:AEC}
Such active cloaks,
including the active interior cloaks proposed earlier by Miller \cite{Miller:2006:PC}, and experimentally tested by Selvanayagam and Eleftheriades \cite{Selvanayagam:2013:EDA},
have the advantage that they are broadband, but the disadvantage that the sources need to be tailored according to the incident field. 
For the plate equation realistic active cloaks have been proposed where the cloak is tailored both according to the incident field and
according to the object one wants to cloak \cite{ONeill:2015:ACI}.

The question arises in cloaking due to anomalous resonance as to what happens when one has two or more cloaking devices (such two or more cylindrical superlenses)
and their ``cloaking regions'' overlap. One might hope that one would get cloaking in the union of the two regions, and this may then help to design cloaks having
cloaking regions of desired shapes. Unfortunately, our results in \cite{McPhedran:2009:CPR} indicate that cloaking is destroyed in the overlap region.

%I find these last two paragraphs a bit too speculative, and to me it is not clear that cloaking due to anomalous resonance creates a quiet zone.
%As has been pointed out already, cloaking by anomalous resonance creates a {\em quiet zone} outside the cloaking system in which objects are hidden from detection. This means it is capable of complementing classical cloaking systems, for which the object is hidden only when inside the cloak. In the latter case, an object approaching the cloak disappears abruptly on reaching its boundary, providing an indication as to the location of the cloak. A system combining both types of cloaking action would then shield both the cloak and objects near it from determination of position information by probes.

%A third aspect is that anomalous resonance cloaks can provide {\em cloaking bricks}, which may be assembled into extended cloaking systems of arbitrary shapes. However, it should be noted \cite{ostrich} that the bricks should be assembled so their cloaking regions come close to each other but do not overlap. The consequence of  anomalous resonance with overlapping cloaking regions is the development of mathematical instabilities \cite{ostrich}.

A challenge for cloaking by anomalous resonance is that it has yet to be experimentally demonstrated. It is hoped that such a demonstration will become available in the near future.

\section{A possible future research direction}
An interesting topic  which may prove a fruitful research direction concerns density of states and related concepts in connection with anomalous resonance. Density of states functions have long been a central tool in solid state physics, and indeed were at the heart of
the 1987 paper by Eli Yablonovitch \cite{Yablonovitch:1987:ISE},
 % ( E. Yablonovitch 1987;PRL 58:2059-2062) 
 which provided one of the two launching pads for the field of photonic crystals. Yablonovitch pointed out that if one could create a structure having band gaps for photons rather than electrons, one would then have in bandgap frequency ranges zero density of state for photons, and thus through Fermi's Golden Rule drastically inhibited spontaneous emission.  The topic of density of states functions  in photonic crystals was then  much discussed, and various density of
state functions were introduced- see for example \cite{McPhedran:2004:DSF}.
 %(McPhedran,R.C., Botten, L.C., McOrist,J., Asatryan, A.A., de
%Sterke,C.M., and Nicorovici, N.A.: {\em Density of states
%functions for photonic crystals} Physical Review E, {\bf
%69},016609  (16 pp.)(2004) ). 

The density of states function itself is only a function of frequency, but one may introduce a local density of states (LDOS) which depends both on frequency and position within a structured material. Note that the spatial integral of the LDOS is related to the classical density of states (DOS)- the spatial integral in the case of a periodic system is just that over the Wigner-Setiz cell. Similarly, there exists a spectral density of states function (SDOS) which depends on frequency and direction of emission. For periodic systems, its integral over the Brillouin zone gives the DOS.   All three of these functions may be constructed 
from the appropriate Green's function by taking its imaginary part at the source point, on, in the case of the SDOS, in the direction of source emission.

Related to these densities of state functions  there are their Hilbert (or causal) transforms, which relate to the frequency shift between the source natural frequency and its emitted  frequency in the structured material. This relates to the difference between the real part of the
Green's function at the source point and its value in a homogeneous reference material, and is called the anomalous Lamb shift- see
\cite{Fussell:2005:DRL}
% Fussell, D.P.,McPhedran, R.C., and de Sterke, C.M.: {\em Decay rate and level shift in a
%circular dielectric waveguide},
%Phys. Rev. A {\bf 71}, 013815 (2005)(15 pp.) 
and \cite{Asatryan:2006:FSS}.
%  Asatryan, A.A., Botten, L.C., Nicorovici, N.A., McPhedran,
%R.C. and de Sterke, C.M.: {\em Frequency Shift of Sources Embedded
%in Finite Two-dimensional Photonic Clusters}, Waves in Random and
%Complex Media, {\bf 16}, 151-165 (2006). 

In the field of metamaterials, density of state concepts have chiefly been of interest in relation to hyperbolic metamaterials, where permittivities and/or permeabilities have different signs in different directions - see \cite{Jacob:2010:EPD}.
 The result of this strong anisotropy is that the photonic density of states  is no longer bounded, but may become very large for particular frequency ranges and directions.
Systems exhibiting anomalous resonance offer similar possibilities and interest. Since the essential singularity at the heart of anomalous resonance is the limit point of a sequence of poles and zeros, this strongly suggests it is associated with an infinite density of states, but only at one frequency. The reasoning here refers immediately to the DOS, but given that the DOS is the spatial integral of the LDOS, flows on to the latter. In fact equation (2.6),
[and to a lesser degree (2.25), (2.35), (2.44), (2.45), (4.27) and (5.23)] in the  paper \cite{Milton:2006:CEA} shows that the imaginary part of the Green's function is
infinite within the cloaking region.  In fact it is infinite in two-dimensions simply for a dipole at any point outside a disk of dielectric constant $-1+i\delta$ surrounded by a medium
of dielectric constant 1, in the limit $\Gd\to 0$. As is well known, this problem can be solved using the method of images and the magnitude of the image dipole blows up as $\delta\to 0$.
Of course these two-dimensional geometries correspond to cylindrical geometries in three dimensions with a line dipole outside, and this is not the same as the point dipole
required for the LDOS. For a point dipole within the cloaking region surrounding the slab lens the quasistatic LDOS is infinite, as can be seen from (5.14) and (5.23) in
\cite{Milton:2006:CEA}. It is even infinite for a dipole at any point outside a sphere of dielectric constant $-1+i\delta$ surrounded by a medium
of dielectric constant 1, in the limit $\Gd\to 0$. This can be seen from the expressions for the image dipole and accompanying
image charge or charge distribution (see \cite{Neumann:1883:HUN} and table 4.2 on page 72 in \cite{Poladian:1990:ETO}) which blow up in magnitude as $\Gd\to 0$.

It would be of great interest to understand in more detail the various density of state functions and their behavior in systems undergoing anomalous resonance. It is tempting to speculate that the
spontaneous emission would happen infinitely fast. But this would imply a mix of frequencies and the infinite LDOS only occurs at one frequency. 

Certain other obvious questions spring to mind:
\begin{enumerate}
\item What are the effects of  frequency, dispersion and loss on densities of state near anomalous resonances?
\item What happens to frequency Lamb shifts near  anomalous resonances?
\end{enumerate}
Those looking through the window offered by densities of states concepts will no doubt find other questions begging answers. However, perhaps the fundamental challenge
remaining is to strengthen the bridge between static and dynamic formulations near an essential singularity of the former.
%%%%%%%%%%%%%%%%%%%%%%%%%%%%%%%%%%%%%%%%%%%%%%%%%%%%%%%%%%%%%%%%%%%%%%%%%
\section*{Acknowledgements}
GWM thanks the National Science Foundation for support through grant DMS-1814854. The referees are thanked for their helpful comments and suggested references. 
%%%%%%%%%%%%%%%%%%%%%%%%%%%%%%%%%%%%%%%%%%%%%%%%%%%%%%%%%%%%%%%%%%%%%%%%%%%%%%%%%%%%%%%%%%

\end{document}